\documentclass[twocolumn]{aa}
\usepackage{graphicx}
\usepackage{txfonts}
\begin{document}

\title{ISOPHOT 95 {\rm $\mu$m} observations in the Lockman Hole} 
\subtitle{The catalogue and an assessment of the source counts\thanks{Based on 
observations obtained with 
the {\sl Infrared Space Observatory}, an ESA science missions with instruments
and contributions funded by ESA Member States and the USA (NASA).}
}

\author{G. Rodighiero
        \inst{1}
        \and 
       A. Franceschini
        \inst{1}
       }

   \offprints{G. Rodighiero}
     
    \institute{
	Dipartimento di Astronomia, Universit\`a di Padova, Vicolo dell'Osservatorio 2,  I-35122 Padova, Italy
        \email{rodighiero@pd.astro.it}
         }
\date{}
   \date{Received ; accepted}

   \abstract{
We report results from a new analysis of a deep 95 {\rm $\mu$m} imaging 
survey with ISOPHOT on board the Infrared Space Observatory, over a $\sim$1 
deg$^2$ area within the Lockman Hole, which extends the statistics
of our previous study (Rodighiero et al. 2003).
Within the survey area we detect sixty-four sources with S/N$>3$ (roughly 
corresponding to a flux limit of 16 {\rm mJy}). Extensive simulations indicate that
the sample is almost complete at $S_{95 \mu m} \geq 100$ {\rm mJy}, while the 
incompleteness can be quantified down to $\sim$30 {\rm mJy}.
The 95 {\rm $\mu$m} galaxy counts reveal a steep slope at $S_{95 \mu m} \le 100$ 
{\rm mJy} ($\alpha\simeq 1.6$), in excess of that expected for a non-evolving source 
population. In agreement with counts data from ISO at 15 and 175 {\rm $\mu$m}, 
this favours a model where the IR populations evolve both
in number and luminosity densities. We finally comment on some differences 
found with other ISO results in this area.

   }

   \maketitle
%

\section{Introduction}

The infrared sky observed from space is providing tight constraints
on the evolution of cosmological source populations.
The IRAS extragalactic number counts (Oliver et al. 1992, Bertin et al. 
1997) showed some marginally 
significant excess of galaxies compared to non-evolutionary predictions.
More recently, ISOPHOT (Lemke et al. 1996) found some evidence of evolution
for the IR population  at 170{\rm $\mu$m} 
(e.g. Puget et al. 1999, Dole et al. 2001),
but source confusion at these wavelengths limits the observations to
moderately faint fluxes ($S_{175\mu m}>135$ {\rm mJy}).
At shorter wavelengths, ISOCAM (Cesarsky et al. 1996) detected a number of faint
sources consistent with strong evolution in the mid-IR  (e.g. Elbaz et al. 1999), 
a factor $\sim$10 in excess with respect to non-evolution predictions.

The ISO deep counts, together with those in the submillimeter 
(e.g. Scott et al. 2002) are used as constraints on models of galaxy
formation and evolution (e.g. Lacey et al. 1994,
Franceschini et al. 2001,  Lagache et al. 2003). 
These studies are also needed to estimate the confusion noise 
for future IR and submillimeter telescopes (Xu et al. 2001, Lagache et al. 2003).

In this paper we present a new analysis of ISOPHOT 95 $\mu$m data in the direction of the 
Lockman Hole, extending our previous study (Rodighiero et al. 2003, hereafter Paper I) and 
significantly improving the source statistics.

The Spitzer Space Observatory (Fazio et al., 1999) will soon observe the same
region of the Lockman Hole in complementary wavebands at 24, 70 and 160 {\rm $\mu$m} with MIPS
and in the near-IR with IRAC. This will provide additional extensive information on the Spectral
Energy Distributions (SEDs) of ISO sources.

The paper is organized as follows: 
in Section 2 we introduce the dataset and in Section 3 we summarize features of 
the adopted reduction procedure.
The infrared maps and the new catalogue are presented in Section 4. The source counts and a
discussion of the effects of confusion and cosmic variance are reported in Section 5.

\section{The sample: LHEX + LHNW}  \label{observation} 

The Lockman Hole (Lockman et al., 1986) is the sky region
with the lowest HI column density, and is thus best suited for the  
detection of faint extragalactic infrared sources. 
Moreover, available multiwavelength observations in this area (Hasinger et al. 2001,
Fadda et al.  2004 in prep., and Rodighiero et al., 2004 in prep.,
Rodighiero et al., 2003, Kawara et al., 2004, Scott et al., 2002,
De Ruiter et al., 1997, and Ciliegi et al. 2003),
allow to study the spectral energy distributions of the detected sources and to address 
their physical nature.
 
ISOPHOT, on board the ISO satellite, observed two different regions in the Lockman Hole. 
Each one of  
the two fields, called LHEX and LHNW, covers an area of  $\sim44'\times 44'$ and has been 
surveyed at two far-infrared wavebands with the C100 and 
C200 detector (respectively at 95 and 175 {\rm $\mu$m}), in the P22 survey raster mode. 
The ISOPHOT survey in the LHEX field has been reported in Paper I, while we
extend here our analysis to the LHNW, which consists of a mosaic    
of four rasters, each one covering an area of $\sim22'\times 22'$. 
The observational parameters are reported in Paper I.

\begin{figure} 
  \centering
   \includegraphics[width=0.5\textwidth]{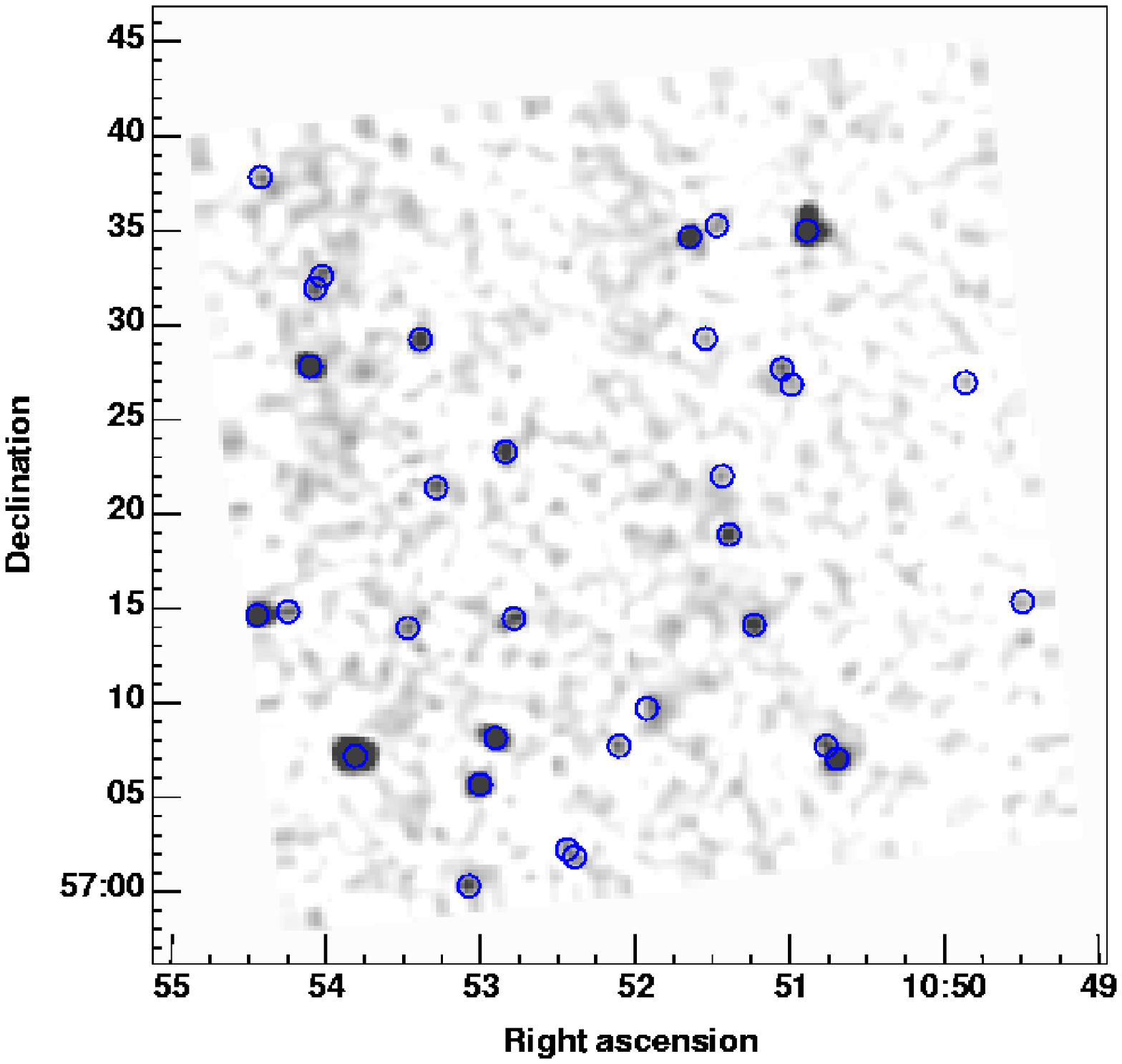}
   \includegraphics[width=0.55\textwidth]{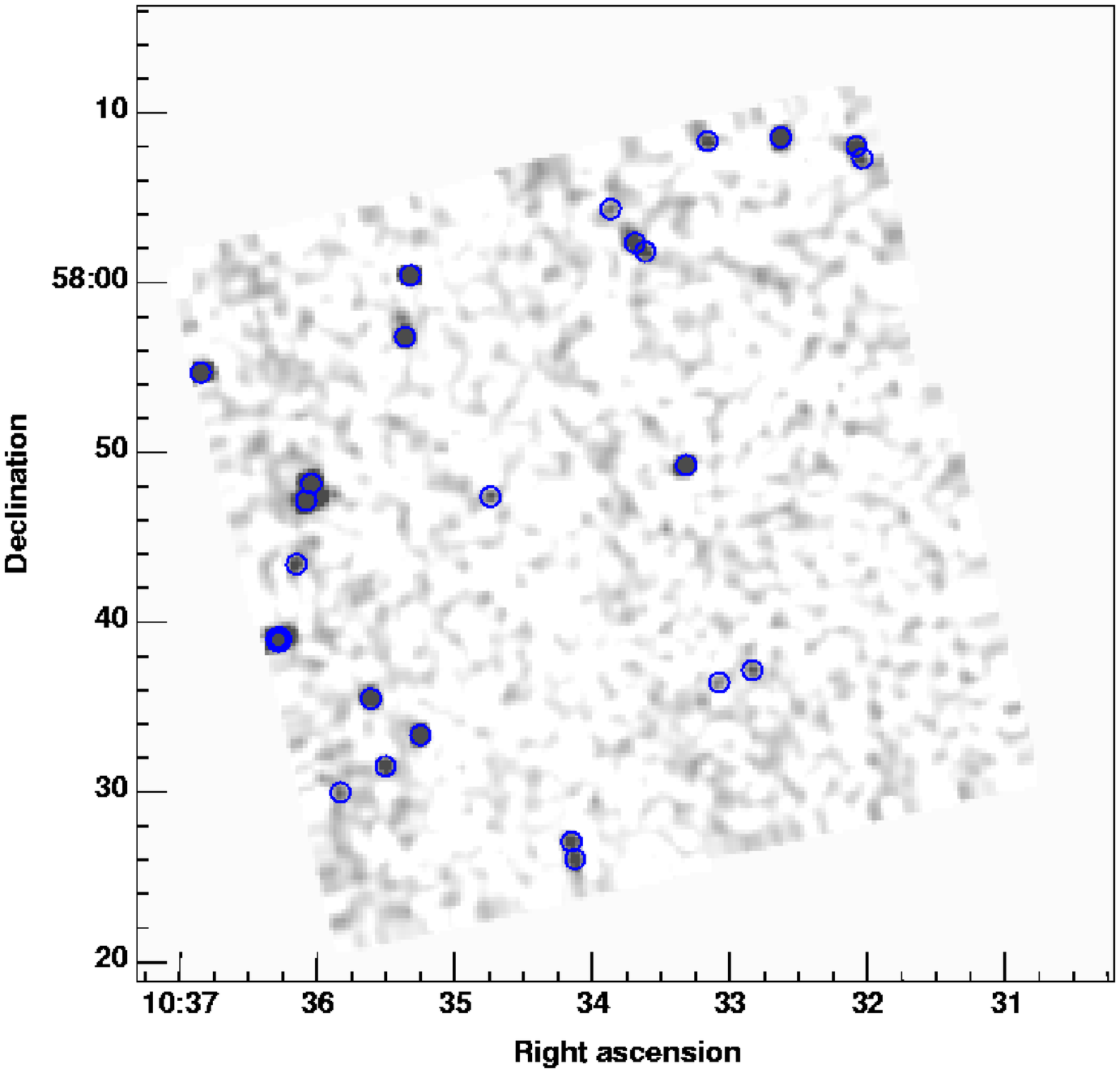}
\caption{The 95 {\rm $\mu$m} map of the Lockman Hole LHEX (upper panel) and LHNW (lower panel) areas. 
The circles indicate our detected sources with signal-to-noise ratio $>$ 4.
Each map is $\sim44'\times 44'$.} 
\label{mosaic} 
\end{figure} 

\section{Data reduction}

The LHNW data set has been reduced with the procedure described in Paper I and 
successfully applied to the LHEX field. 
The procedure is based on a parametric algorithm that fits the signal time
history of each detector pixel. It extracts the background
level, in order to identify the singularities induced by cosmic-rays impacts
and by transient effects in the detectors, and to identify and extract real sky sources.

The source extraction process, the completeness, the photometric and
astrometric accuracies of the final catalogues have been tested by us with 
extensive sets of simulations, by inserting into the real image sources
with known flux and position, and accounting for all the details of the 
reduction procedure. We defer the interested reader to Paper I for details.
The flux calibration was also verified by reducing with the
same technique ISOPHOT 95 $\mu$m  observations of some selected standars,
such as calibrating stars and IRAS sources, in order to cover the flux range
spanned by our Lockman source catalogue.

\section{The 95 {\rm $\mu$m} source catalogue}
\label{catalog_sec} 

In Paper I, we detected 36 sources with S/N$>3$ (corresponding 
to a flux of 16 {\rm mJy}) within the LHEX area, making up a complete
flux-limited sample for $S_{95 \mu m} \geq 100$ {\rm mJy}. Reliable
sources were detected, with decreasing but well-controlled completeness,
down to $S_{95 \mu m} \simeq 20$ {\rm mJy}.
In the twin LHNW area of $\sim$0.5 deg$^2$, we detect 28 sources above 
a 3-$\sigma$ threshold, reaching the same flux level as in the previous sample.

In Figure \ref{mosaic} we show the final maps obtained in the LHEX and in the LHNW
fields. The open circles indicate  sources detected with a signal-to-noise ratio 
greater than 4. 
In Table \ref{catalog} we report the new list of sources detected in the LHNW 
region. The entries are: IAU-conformal names, sky coordinates 
(right ascension and declination at 
Equinox J2000), the detection significance (signal-to-noise ratio), 
the 95 {\rm $\mu$m} total fluxes (in {\rm mJy}) and their uncertainties. 
As discussed in Paper I, all sources have been extracted from the 
maps and confirmed by visual inspection of the pixel history. This approach 
produces an highly reliable source catalogue. 

\begin{figure} 
  \centering
   \includegraphics[width=0.3\textwidth]{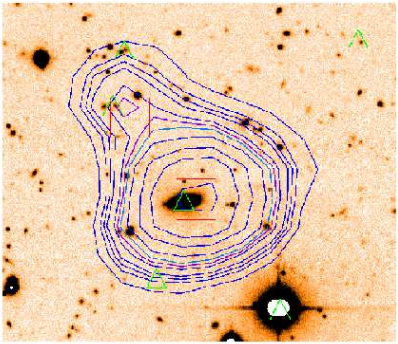}
 \caption{Example of a pair of ISOPHOT complex sources. The Figure shows the optical
 R band image
(Fadda et al. 2004, in prep.) with overlayed the 95 $\mu$m contours. The ISOPHOT
positions are marked with open squares, the triangles indicate the ISOCAM 15 {\rm $\mu$m} 
detections (Fadda et al., 2004).
}
 \label{ids} 
 \end{figure}

Our source extraction tool and simulation procedures can be optimally used to 
recover the total 
fluxes for slightly blended sources. 
An example is reported in Figure \ref{ids}, where for a pair of ISOPHOT complex sources
(sources 8 and 17 in LHEX, see catalogue in Paper I) we show the optical R band image
(Fadda et al. 2004, in prep.) with overlayed the 95 {\rm $\mu$m} contours. The ISOPHOT
positions are marked with open squares, the triangles indicate the ISOCAM 15 {\rm $\mu$m} detections
(Fadda et al., 2004). The distance between the two far-IR sources, $\sim55$ arcseconds,
typical for the few blended objects in the LHEX and LHNW maps, slightly exceeds
the ISOPHOT C100 beam ($\sim$ 45 arcsec).

The final combined LHEX+LHNW catalogue contains a total of 64 sources,
detected at 95 {\rm $\mu$m} over an area of about 1 square degree.

\begin{center}
\begin{table} 
\scriptsize
\centering
\caption{LHNW source catalogue.}
\begin{tabular}{|l|c|c|c|c|c|} 
\hline 
\hline 
ID& RA     &  DEC    & S/N & Flux  \\ 
~ &(J2000) & (J2000) &~    & [{\rm mJy}] \\
\hline 
\hline   
LHJ103521+580034  &10:35:21.2  &     +58:00:34  &  33  &   235 $\pm$  24\\
LHJ103606+574715  &10:36:06.7  &     +57:47:15  &  22  &   163 $\pm$  17\\
LHJ103604+574815  &10:36:04.4  &     +57:48:15  &  21  &   181 $\pm$  20\\
LHJ103653+575444  &10:36:53.8  &     +57:54:44  &  16  &   201 $\pm$  24\\
LHJ103618+573904  &10:36:18.2  &     +57:39:04  &  16  &   227 $\pm$  26\\ 
LHJ103515+573330  &10:35:15.5  &     +57:33:30  &  16  &   136 $\pm$  16\\
LHJ103523+575656  &10:35:23.3  &     +57:56:56  &  15  &   114 $\pm$  14\\
LHJ103318+574925  &10:33:18.6  &     +57:49:25  &  15  &   159 $\pm$  19\\
LHJ103236+588042  &10:32:36.2  &     +58:80:42  &  12  &   109 $\pm$  14\\
LHJ103537+573537  &10:35:37.4  &     +57:35:37  &  12  &   114 $\pm$  13\\
LHJ103530+573139  &10:35:30.8  &     +57:31:39  &   8  &    73 $\pm$  12\\
LHJ103409+572715  &10:34:09.1  &     +57:27:15  &   8  &    63 $\pm$  10\\
LHJ103407+572613  &10:34:07.5  &     +57:26:13  &   8  &    69 $\pm$  10\\ 
LHJ103341+582031  &10:33:41.2  &     +58:20:31  &   8  &    77 $\pm$  12\\
LHJ103202+588009  &10:32:02.4  &     +58:80:09  &   8  &    97 $\pm$  15\\
LHJ103610+574330  &10:36:10.6  &     +57:43:30  &   7  &    85 $\pm$  14\\
LHJ103336+582000  &10:33:36.7  &     +58:20:00  &   6  &    34 $\pm$  7\\
LHJ103308+588030  &10:33:08.8  &     +58:80:30  &   5  &    80 $\pm$  16\\
LHJ103445+574733  &10:34:45.1  &     +57:47:33  &   5  &    48 $\pm$  10\\ 
LHJ103249+573719  &10:32:49.7  &     +57:37:19  &   5  &    56 $\pm$  11\\ 
LHJ103352+584030  &10:33:52.1  &     +58:40:30  &   4  &    55 $\pm$  14\\
LHJ103304+573637  &10:33:04.3  &     +57:36:37  &   4  &    55 $\pm$  14\\ 
LHJ103550+573005  &10:35:50.4  &     +57:30:05  &   4  &    49 $\pm$  11\\
LHJ103159+587026  &10:31:59.8  &     +58:70:26  &   4  &   100 $\pm$  24\\
LHJ103524+575749  &10:35:24.5  &     +57:57:49  &   3  &    35 $\pm$  7\\
LHJ103316+573136  &10:33:16.1  &     +57:31:36  &   3  &    30 $\pm$  10\\
LHJ103327+574539  &10:33:27.6  &     +57:45:39  &   3  &    42 $\pm$  12\\
LHJ103313+589010  &10:33:13.2  &     +58:90:10  &   3  &    21 $\pm$  7\\
\hline 
\hline 
\end{tabular} 
\label{catalog} 
\normalsize
\end{table} 
\end{center}

\section{Source counts from the Lockman Hole 95 {\rm $\mu$m} survey} 
\label{countslh}  

We have computed the 95 {\rm $\mu$m} source counts down to 
a flux level of 30 {\rm mJy}, following the same procedure described in Paper I. 
The counts have been corrected for the incompleteness factors
derived through simulations. According to these, the completeness decreases 
from $\sim 80\%$ around 100 {\rm mJy}, to $\sim 30\%$ around 40 {\rm mJy}, see Table 1 in Paper I).
Poissonian noise has been considered.

\begin{figure} 
  \centering
   \includegraphics[width=0.5\textwidth]{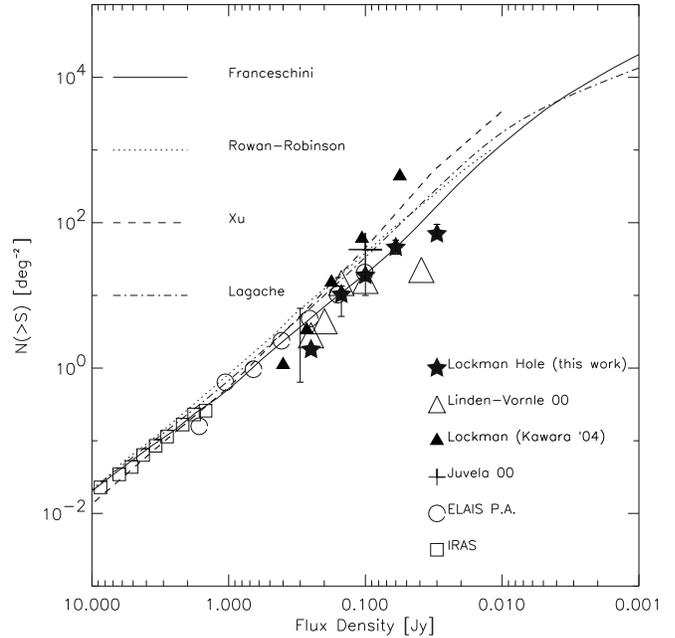}
 \caption{Integral counts at 95 {\rm $\mu$m}. Our estimated values 
(starred filled symbols) are compared with data from a variety of other surveys 
(see text for references and details). The counts are expected to be dominated 
by extragalactic sources, for which some model predictions are also reported.
}
 \label{counts_i} 
 \end{figure}

The new source counts confirm our previous finding for the LHEX field.
The integral counts are plotted in Figure \ref{counts_i} as starred symbols.  
Our results are compared with those from other surveys: the preliminary 
analysis of the ISOPHOT ELAIS survey (Efsthatiou et al. 2000, open circles);
the counts derived from the IRAS 100{\rm $\mu$m} survey (open squares); 
those published by Linden-Vornle et al. (2000, open triangles),
Juvela et al. (2000, '+' symbol) and recently by Kawara et al. (2004, filled triangles) which
are based on the same dataset (see discussion below).
A comparison is made also with model counts by Franceschini et al. (2001, solid line), 
Lagache et al. (2002, dot-dashed line), Xu et al. (2001, dashed line)
 and Rowan-Robinson et al. (2001, dotted line).
The slope of the counts turns out to be $\alpha \sim 1.6$.

The good agreement with the ELAIS source counts (which are based on a 
larger area) suggests that these results are not much affected by spatial biases.
In the fainter flux region below 100 {\rm mJy}, the source counts
from different surveys appear instead more scattered.

\begin{figure} 
\centering
   \includegraphics[width=0.5\textwidth]{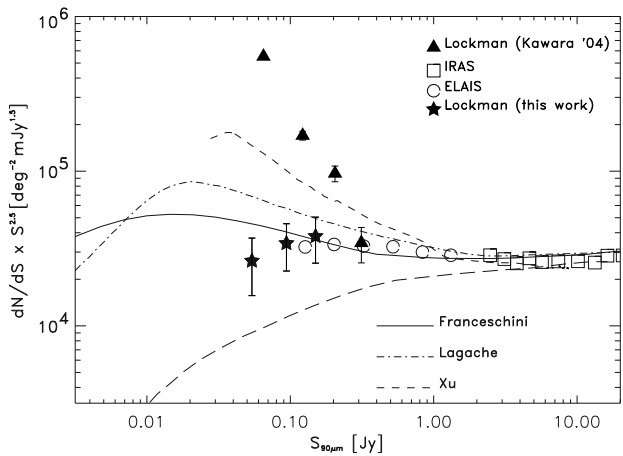}
\caption{Differential 95 {\rm $\mu$m} counts dN/dS normalized to the Euclidean law
($N \propto S^{-2.5}$). Symbols are the same as in Fig. \ref{counts_i}.
Our results are compared with those from the preliminary 
analysis of the ISOPHOT ELAIS survey (Efsthatiou et al. 2000, open circles)
and  with those from Kawara et al. (2004, filled triangles).
We also show as a long-dashed line the expected contribution of non-evolving
spirals as in the model of Franceschini et al. (2001). 
} 
\label{counts_d} 
\end{figure}

In Figure \ref{counts_d} we report the differential 95 {\rm $\mu$m} counts dN/dS
normalized to the Euclidean law $N\propto S^{-2.5}$
(the data are reported in Table \ref{ci}).  For this we have used only sources with 
a signal-to-noise ratio greater than 4.
The results are compared with those from Kawara et al. (2004, filled triangles)
and from the preliminary analysis of the ISOPHOT ELAIS survey (Efsthatiou et al. 2000, open circles).
A comparison is performed with modellistic differential counts from Franceschini et al. 
(2001), Xu et al. (2001) and Lagache et al. (2002).

Our results differ from those of Kawara et al. (2004),
who analysed the same dataset used here (i.e. the combined 
LHEX and LHNW ISOPHOT 95 {\rm $\mu$m} data), but with an independent reduction 
based on the standard PHOT Interactively Analysis (PIA, Gabriel et al. 1997). 
Their final catalogue includes quite more sources than ours (223 versus 64),
at similar flux levels and S/N ratios.
The two number counts turn out to be in good agreement only above 200 {\rm mJy}. 
Although unable to understand the origin of this difference, we note that different
flux calibrations were adopted by the two analyses, 
Kawara et al. using a single reference, IRAS F10507+5723.
In any case, our careful detector time-sequence analysis and the simulations 
performed to check each sources should in principle allow us a robust rejection 
of spurious detections, while at the same time keeping under control the survey 
incompleteness.

With this new extended sample we confirm our conclusions in Paper I about
the existence of an evolving population of IR galaxies, consistent with
independent results based on deep ISOCAM mid-IR and ISOPHOT 175 {\rm $\mu$m} counts. 
Our analysis, however, indicates that the evolution cannot be too strong since the 
level of the counts at the faintest fluxes is not very high.
The counts predictions reported in the figures typically assume that the 
whole local galaxy population evolves back in cosmic time in luminosity and/or 
number density. With such an assumption, the model counts tend to more or 
less exceed our data at the faint fluxes. In the evolutionary scheme by 
Franceschini et al. (continuous lines) the contribution of a non-evolving 
normal galaxy population dominating the bright counts adds to that of a
fast evolving starburst-galaxy component rising up at faint fluxes, then producing 
relatively flat overall counts.
The present results at 95 {\rm $\mu$m} seem to confirm this model.

An apparently non-random spatial distribution of the 95 {\rm $\mu$m} sources 
was already indicated in Paper I. 
A comparison of the two maps in LHEX and LHNW (Fig. \ref{mosaic}) 
shows even more clearly the effects of field-to-field variation (28 versus 36
sources detected in the to areas).
The source spatial distribution in LHNW seems particularly inhomogeneous,
with source clustering close to two map borders.
This strong clustering of the far-IR sources (whose physical origin may be
related with the effects of galaxy interactions and mergers as a trigger
of the starburst-induced far-IR emission) makes the confusion problem
very serious. Four sources in LHNW and five in LHEX appear with close
companions (see one example in Fig.\ref{ids}), emphasizing that source 
blending and confusion happens even at these moderately faint flux limits.
Although progress is being achieved with Spitzer, 
definitely larger space telescopes (e.g. the Herschel Space Observatory,
due to launch in 2007) will be required for proper sampling of this
difficult waveband domain.

\begin{center}
\begin{table} 
\caption{Integral and differential source counts at 95 {\rm $\mu$m}.} 
\small
\begin{tabular}{l l  c  c   c c c} 
\cline{1-2}
\cline{4-7} 
\multicolumn{2}{c}{Integral}&&\multicolumn{3}{c}{Differential}\\
\cline{1-2}
\cline{4-7} 
S  & dN($>$S) 	&&	flux bin &bin center & $dN/dS \times S^{2.5}$ \\
{\rm mJy}   & deg$^{-2}$&&	{\rm mJy}      & {\rm mJy}       & deg$^{-2}${\rm mJy}$^{1.5}$   \\
\cline{1-2}
\cline{4-7} 
30    & 70.0 & & 30-100   & 54  & $2.63e4\pm 1.06e4$\\
60    & 45.7 & & 60-150   & 94  & $3.42e4\pm 1.12e4$\\
100   & 18.8 & & 100-300  & 173 & $3.79e4\pm 1.25e4$\\
\cline{4-7}
150   & 10.3 & & & & & \\
300   & 1.8  & & & & &\\
\cline{1-2}
\end{tabular} 
\label{ci} 
\normalsize
\end{table} 
\end{center}

\section*{Acknowledgments}
 
We acknowledge many exchanges and support from C. Lari.
We thank an anonymous referee.
This work was partly supported by the "POE" EC TMR Network Programme 
(HPRN-CT-2000-00138).

\end{document}